\documentclass[useAMS,usenatbib,usegraphicx]{mn2e}

\usepackage{color}

\title[Emission of spreading layer in neutron stars]{On the spreading layer emission  in luminous accreting neutron stars}
\author[M.G. Revnivtsev, V. F. Suleimanov  and J. Poutanen]{Mikhail G. Revnivtsev$^{1}$\thanks{E-mail: revnivtsev@iki.rssi.ru}, 
Valery F. Suleimanov$^{2,3}$ and Juri Poutanen$^{4}$\\
$^{1}$ Space Research Institute, Russian Academy of Sciences, Profsoyuznaya 84/32, 117997 Moscow, Russia\\
$^{2}$ Institut f\"ur Astronomie und Astrophysik, Kepler Center for Astro and Particle Physics,\\
     Universitaet T\"ubingen, Sand 1, D-72076 T\"ubingen, Germany\\
$^{3}$ Kazan (Volga region) Federal University, Kremlevskaya str., 18, Kazan 420008, Russia\\
$^{4}$ Astronomy Division, Department of Physics, PO Box 3000, FI-90014 University of Oulu, Finland\\
}
\begin{document}


\date{\today}
\pagerange{\pageref{firstpage}--\pageref{lastpage}} \pubyear{2013}

\maketitle

\label{firstpage}

\begin{abstract}
Emission of the neutron star surface potentially contains information about its size and thus of vital importance for high energy astrophysics. 
In spite of the wealth of data on the emission of luminous accreting neutron stars, the emission of their surfaces is hard to disentangle from their time averaged spectra. 
A recent X-ray transient source XTE J1701$-$462 has provided a unique dataset covering the largest ever observed luminosity range 
for a single source and showing type I (thermonuclear) X-ray bursts. 
In this paper, we extract the spectrum of the neutron star surface (more specifically, the spectrum of the boundary layer between the inner 
part of the accretion disc and the neutron star surface) 
with the help of maximally spectral model-independent method. 
We show compelling evidences that the energy spectrum of the boundary layer stays virtually the same over factor of 20 variations of the source luminosity. 
It is rather wide and cannot be described by a single temperature blackbody spectrum,  
probably because of the inhomogeneity of the boundary layer and a spread in the colour temperature. 
The observed maximum colour temperature of the boundary/spreading layer emission of $kT\approx$2.4--2.6 keV  
is very close to the maximum observed colour temperature in the photospheric radius expansion X-ray bursts,
which is set by the limiting Eddington flux at the neutron star surface. 
The observed stability of the boundary layer spectrum and its maximum colour temperature strongly supports 
theoretical models of the 
boundary/spreading layers on surfaces of luminous accreting neutron stars, which assume the presence of a region emitting at the local Eddington limit. 
Variations in the luminosity in that case lead to changes in the size of this region, 
but affect less the spectral shape.
Elaboration of this model will provide solid theoretical grounds for measurements of the neutron star sizes 
using the emission of the boundary/spreading layers of luminous accreting neutron stars.
\end{abstract}


\section{Introduction}

Compact objects in binary systems accreting matter from their binary components reveal themselves as bright X-ray emitters.
Matter gradually moves closer to the central compact object and is heated  to tens of millions degrees. 
The emergent X-ray radiation potentially contains information about the compact object and 
about the behaviour of matter in strong gravitational and magnetic fields.
Thus, it is necessary to improve our knowledge about formation of the X-ray emission in such accreting sources 
before we are able to extract parameters of compact objects such as their radii and masses.  
In this paper, we concentrate on emission of binaries which harbour neutron stars  (NSs) as compact objects.

The majority of known bright X-ray sources in our Galaxy are NS binaries. Their X-ray emission was discovered at the dawn of X-ray astronomy. Already first observations of the brightest NS binaries have shown that their emission is likely thermal \citep{chodil68,toor70}. Numerous observations of a set of NS binaries in our Galaxy have shown that the whole variety of their energy spectra can be broadly separated into two main classes, with the spectral cutoff at energies below 6--10 keV and around 50--100 keV, which  are called soft and hard spectral states, respectively 
(here we consider only active sources, which have luminosities $L_{\rm x}>10^{35}$ erg s$^{-1}$). 
Accreting NSs typically demonstrate soft spectra if their luminosities are above $\sim 10^{36}$ erg s$^{-1}$ and hard spectra if their luminosities are lower than that. Simple physical arguments indicate that the soft state spectra form in the optically thick,
while the hard state spectra in the optically thin media \citep[see e.g.][]{B00}.

Accreting low-magnetic field NSs should generate X-ray radiation in at least two geometrically distinct regions: in the accretion disc (similarly to the case of accreting black holes) and in the boundary/spreading layer (BL/SL) between the accretion disc, whose inner parts rotate around the NS with very high velocity, and the NS surface. 
The energy release in these parts of the flow is comparable \citep{SS86,SS00}. 
In the optically thick regime, the  effective temperatures  $T_{\rm eff}$ of these two regions should be approximately 1--2.5 keV. 
The observed colour temperatures are increased by the hardness (colour correction) factor $f_{\rm c}$ 
\citep*{london86,lapidus86,ST95} and decreased by the gravitational redshift $z$ at the NS surface $T_{\rm c} = f_{\rm c} T_{\rm eff}/(1+z)$. 
The contribution of the neutron star surface not covered by the BL/SL is small. 
The usual temperatures of NSs in low mass X-ray binaries in quiescent states are about 0.1--0.3  keV \citep[see e.g.][]{wd13}. 
In the active state, the temperature might be higher, but the large extent and the high temperature 
of the accretion disc and the BL/SL make them dominating the luminosity of the NSs, certainly in the 
high energy band 3--20 keV. 

\begin{figure}
\includegraphics[width=0.95\columnwidth,bb=32 182 570 700,clip]{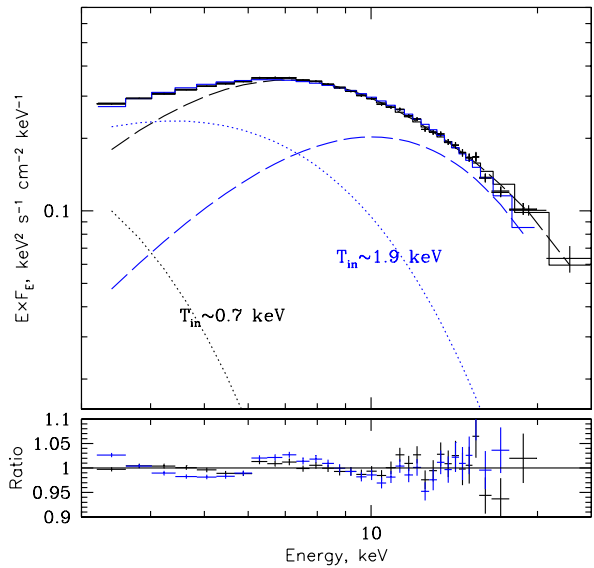}
\caption{Example of decomposition of a typical spectrum of the NS binary in the soft state into the accretion disc and the BL components. 
In two cases we assumed two different shapes of the BL emission component. 
In the first case (shown by blue curves), we took the BL component as a single temperature black body with $kT_{\rm bb}=2.6$ keV.
In the second case (shown by black curves), the BL component was approximated by a sum of two black bodies with 
temperatures of 1.6 and 3.1 keV. 
The dashed curves show the assumed contribution of the BL, and the 
dotted curves are the contribution of the accretion disc (model {\sc diskbb}) with the inner disc temperature $kT_{\rm in}=1.9$  and 
0.7 keV, for the two cases, respectively.  
It can be seen than in spite of quite similar quality of fits (data/model ratios are within $\sim$2\%), the resulting spectral decomposition is drastically different. 
}
\label{fig:example}
\end{figure}

From the observational point of view, spectra of NSs in their soft state are smooth 
and cannot be easily unambiguously decomposed into these anticipated components. 
{\it Virtually regardless of the statistical quality and the energy resolution of the data, a variety of spectral models can be fitted to the spectra with the comparable fit quality. 
This makes the approach based only on a $\chi^2$ fitting technique not persuasive.} 
This ambiguity led to the completely different interpretations of the soft state spectra \citep*{mitsuda84,white88,disalvo02}. 
To justify this statement we show an example of such completely different spectral 
decomposition with the equally good $\chi^2$ (see Fig.~\ref{fig:example} and further discussion in Sect.~\ref{sect:decomp}).

Completely different approach to the problem of spectral decomposition was proposed by \citet{mitsuda84} and elaborated in works of \citet*{gilfanov03} and \citet{revnivtsev06}. It was shown that secure decomposition of the NS spectra can be done with the help of model-independent analysis of their timing variability. 
They demonstrated that flux variations at the time scales smaller than $\sim$1 s  are primarily caused by variations of the BL flux only. Its spectral shape remains nearly constant, while the normalization varies. 
At the same time scales, the accretion disc flux and its spectral shape remain virtually constant (similarly to the 
behaviour of accretion discs around black holes, see e.g. \citealt{churazov01}) thus providing us with a possibility to make model-independent spectral decomposition.

One of the main results obtained with this technique \citep{gilfanov03,revnivtsev06} is that the spectrum of the BL changes very little while its luminosity varies by more than an order of magnitude. This is an important property, which was previously predicted for the BL/SL at the surface of high luminosity NSs by \citet{IS99}. They showed that the matter which is continuously coming from the accretion disc can decelerate and settle to the NS surface only via a layer 
spreading over some part of the surface. The larger the mass accretion rate, the larger the part of the NS surface occupied by the SL. 
It was also shown that the maximum radiation flux 
is always close to the local Eddington flux in a wide range of the luminosities. Optically thick regime of the layer emission means that its spectral shape should be close to that of a diluted blackbody with the maximal colour temperature mainly governed by the NS gravity. 
Variation of the total mass accretion rate in the SL affects only its latitude extent, but not the maximal colour temperature, as it is set by the local Eddington flux \citep{IS99,SulP06}. First attempts to use this property to make estimates of the NS radii were done by \citet{revnivtsev06} and \citet{SulP06}.

A rising pile of the observational data on NS emission and especially the outburst of a unique transient source XTE J1701$-$462, which demonstrated all types of spectral variability patterns previously observed only in different objects, have triggered new attempts to obtain spectral decomposition of the NS spectra based only on  the quality of the spectral fits \citep*{lin07,lin09a,ding11}. By fitting the multicomponent models to the data, these authors obtained that the temperature of the blackbody component, which they ascribe to the NS surface, strongly varies with the source luminosity. 

In the present paper, we  study the emission of the BL/SL of the NS in the X-ray transient XTE J1701$-$462 and   show that it has nearly constant spectral shape when the source  luminosity was above $L_{\rm x}\sim 10^{37}$ erg  s$^{-1}$. These findings strongly support the physical model of the SL of \citet{IS99}.

\begin{figure}

\includegraphics[width=\columnwidth,bb=32 162 600 720,clip]{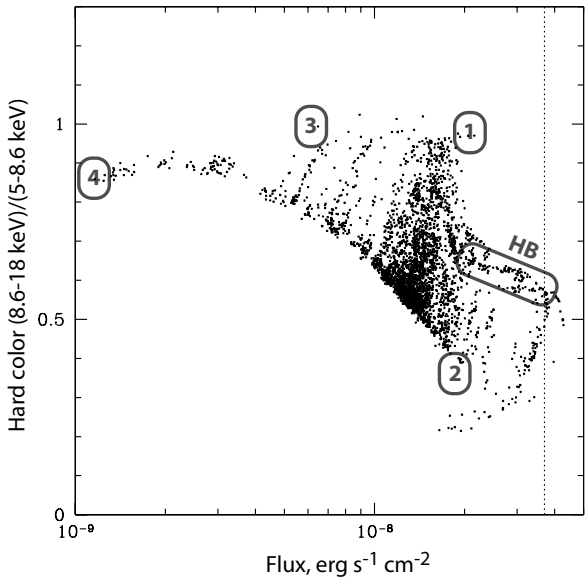}
\caption{The colour -- bolometric flux  diagram for the transient XTE J1701$-$462 calculated over time periods when the source was in its soft spectral states. 
Each point corresponds to 512 s of the integration time. 
The source fluxes in all bands were computed in physical units taking into account all possible long term variations in the energy response of the instrument. 
The dotted vertical line denotes the flux of the source at the photospheric radius expansion phase 
during its thermonuclear (type I) X-ray burst.
}
\label{fig:cid}
\end{figure}

\section{Used datasets}

In this paper, we use numerous observations of the {\it Rossi X-ray Timing Explorer (RXTE)} observatory \citep*{BRW93} of the transient XTE J1701$-$462, performed in the period from 2006 Jan 19 (MJD 53754.78)  till 2007 July 30 (MJD 54311.21). All data were analysed with the help of standard tasks of {\sc heasoft} version 6.8. 
The instrumental background of the Proportional Counter Array (PCA) spectrometer 
aboard of the  {\it RXTE} was estimated using the model {\sc cmbright}, developed by Craig Markwardt \citep{jahoda06}.

Spectral modelling was performed with the help of {\sc xspec} package \citep{Arn96}. All spectral models were multiplied by a model of interstellar absorption {\sc wabs}  with the hydrogen column density $N_{\rm H}=2\times10^{22}$ cm$^{-2}$ \citep{lin09a,fridriksson10}.
During the outburst the source flux varies by a factor of more than 100. In this paper we study only observations, in which the source was in the soft spectral state (MJD 53754.78-54311.21). During these observations, the source showed variations by a factor of $\sim$40 and significant changes in the spectral hardness. The colour--intensity diagram is shown in Fig.~\ref{fig:cid}. 

\section{Spectral decomposition difficulties}
\label{sect:decomp}

The spectral shape of the optically thick accretion disc \citep[see e.g.][]{SS73} in an X-ray source has been 
recognised quite long time ago (see e.g. \citealt*{SLE76}, 
based on observational data of \citealt{tananbaum72}), 
and since then this spectral component has been extensively studied by different authors with the help of different datasets \citep*[see e.g.][for a review]{done07}. 
The black hole accretion discs are simpler in some respect, because luminous black holes typically have only one optically thick region, whose emission can be relatively unambiguously identified in their broad-band spectra. 

The case of luminous NSs is more complicated. While the shape of the optically thick accretion discs around NS might be somehow scaled from those around black holes (however some differences should exist due to the different boundary conditions at the inner boundary of the disc in the two cases), the spectrum of the optically thick BL can be nontrivial.

Typically it is assumed that the BL/SL emits as a black body with the single temperature \citep{mitsuda84,lin07,lin09a,ding11,church12}.  However, what is very important is that the assumed shape of the BL spectrum has a major consequence for the resulting spectral decomposition. 
As an example of such ambiguousness, we present two types of decomposition of the spectrum of the X-ray transient XTE J1701$-$462 in its soft, so called ``atoll'' state using two different assumptions about the shape of the BL spectrum 
(see Fig.~\ref{fig:example}).
In one case, we assumed that the BL spectrum can be approximated by a perfect single-temperature black body ($kT_{\rm bb}\sim2.6$ keV), while in another case  the BL spectrum is approximated by a sum of two black bodies  with  temperatures of 1.6 and 3.1 keV. 
In both cases, we adopted a simplest multicolour blackbody approximation ({\sc diskbb} model in {\sc xspec}  package) for the spectrum of the accretion disc. It is clearly seen that resulting decompositions are drastically different. In these two cases the temperature of the multicolour disc component differs by a factor of more than 2 and the contribution of the accretion disc to the total luminosity differs by almost an order of magnitude (the fainter disc component has a smaller inner temperature $T_{\rm in}$). 
Statistically, both decompositions have good quality, which is essentially limited by systematic uncertainties of the instrument response calibration.  The reduced $\chi^2_{\nu}$ is  $1.02$ for the first case
($kT_{\rm bb}=2.6$ keV, $kT_{\rm in}=1.9$ keV) and  $0.45$ for the second case
($kT_{\rm bb}=1.6$ and $3.1$ keV, $kT_{\rm in}=0.7$ keV). Here are adopted 2\% systematical uncertainties in flux measurements in all energy channels and added them quadratically to pure statistical uncertainties, provided by Poisson noise in these energy channels.

{\it There are no physical arguments why emission of the BL should have the shape of a perfect single temperature black body}. Deviations from pure blackbody or some non-uniformity of blackbody colour temperatures of the BL emission cannot be excluded. Thus,  we should conclude that model-dependent spectral decompositions are inevitably ambiguous. Application of additional ``desirability'' criteria to the model dependent decomposition force the spectral model fitting results behave the way, externally introduced by a researcher, and therefore cannot be considered as a robust solution of this problem. Maximally possible model-independent methods should be employed.

\begin{figure}
\includegraphics[width=0.95\columnwidth,bb=32 182 570 720,clip]{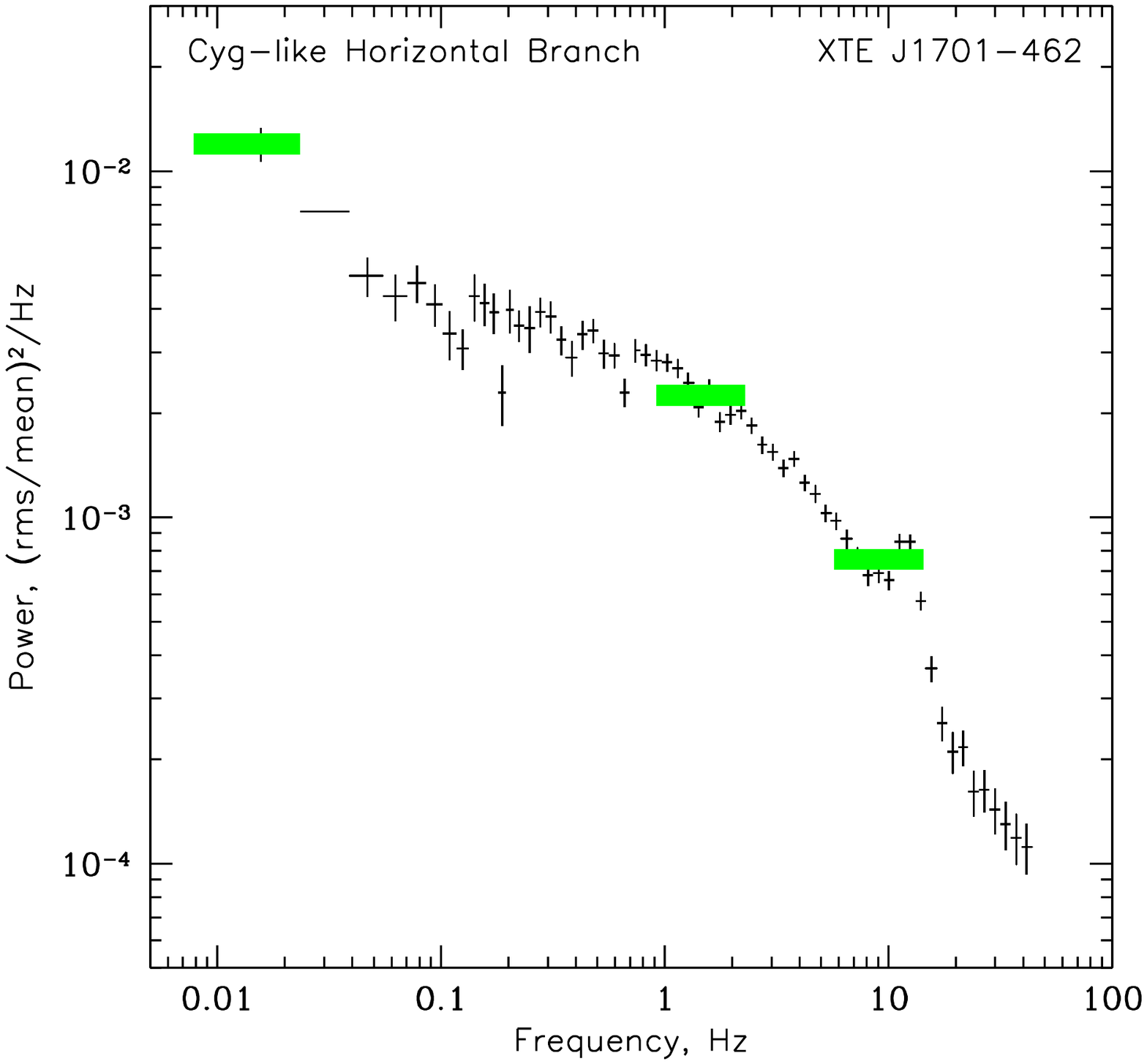}
\includegraphics[width=0.95\columnwidth,bb=32 182 570 720,clip]{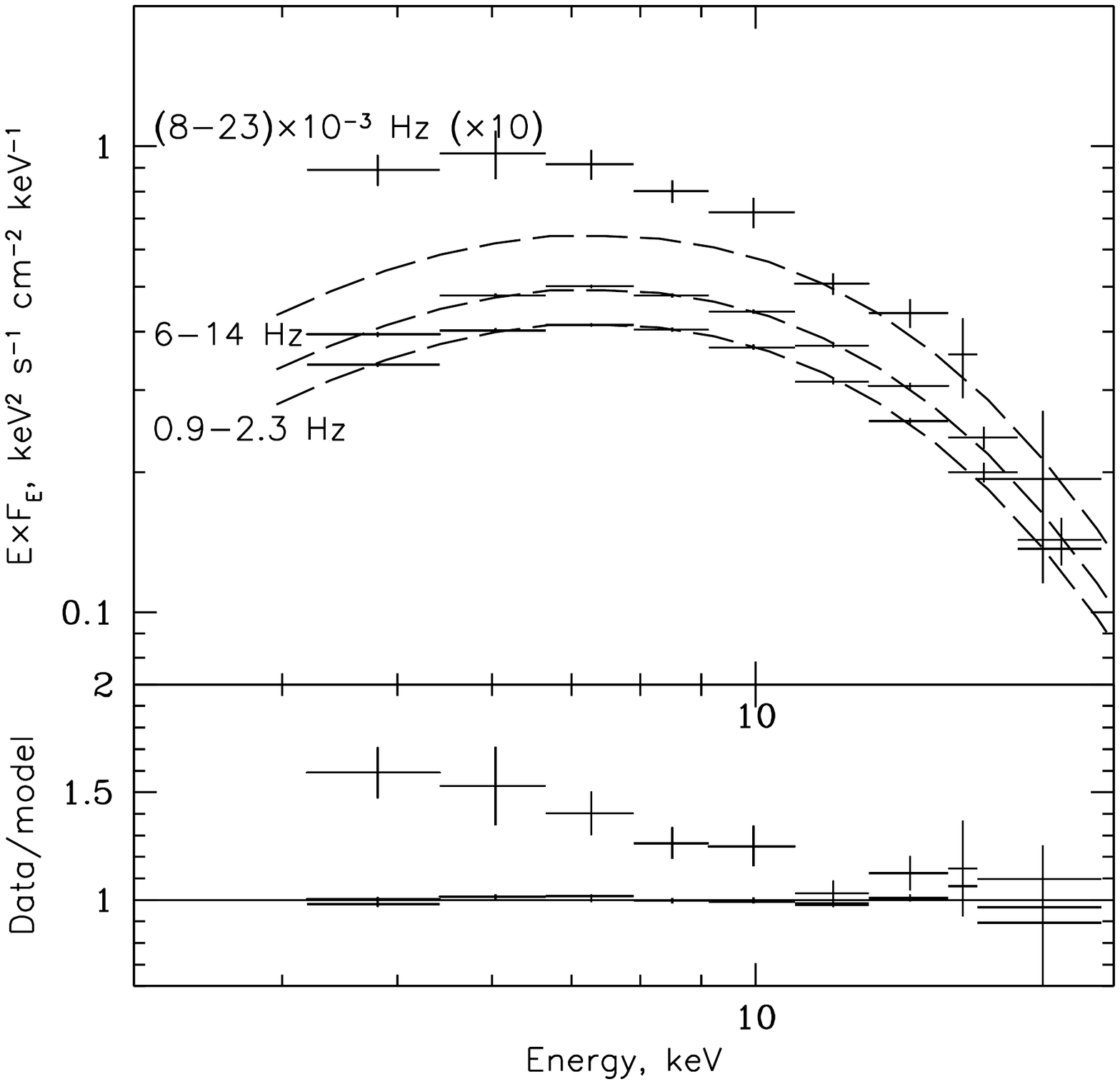}
\caption{Top: The power spectrum of the source flux variability, collected over period MJD 53754.77--53778.93, which covered the so-called horizontal branch on the colour-intensity diagram. The horizontal green
bars show the frequency intervals, for which we show frequency resolved spectra at the bottom panel. 
Bottom: The examples of the 
frequency resolved energy spectra of  XTE~J1701$-$462 at different Fourier frequencies. 
At frequencies above $\sim1$~Hz, the frequency resolved energy spectra have the same spectral shape given 
by equation~(\ref{eq:sl_spectrum}). 
}
\label{fig:freqspectra}
\end{figure}

\section{Spectrum of the BL/SL}
\subsection{Frequency resolved energy spectra}

If one spectral component varies at some time scale (Fourier frequency)  much more than another and its spectral shape does not vary with flux, then we can use these properties to make model-independent spectral decomposition. 
Applicability of this method to the case of luminous NSs was demonstrated by \citet{mitsuda84}, \citet{gilfanov03} and \citet{revnivtsev06}. They showed that the BL/SL component varies at high frequencies much more than the accretion disc (note  that emission of the accretion disc around black holes is also the least variable part of their spectra, see \citealt{churazov01}).
Let us look at the Fourier frequency resolved energy spectra of XTE~J1701$-$462 
(see details of  this techniques in e.g. \citealt*{revnivtsev99}).

The source XTE J1701$-$462 has demonstrated a variety of spectral and timing behaviour. In the soft spectral state, the strongest  variability was observed during the so called ``horizontal'' branch  of the colour-intensity diagram \citep{homan07}. Therefore, we have analysed the data collected during this period, more specifically, during the period MJD 53754.77--53778.93.

\begin{figure}
\includegraphics[width=0.95\columnwidth,bb=32 182 570 720,clip]{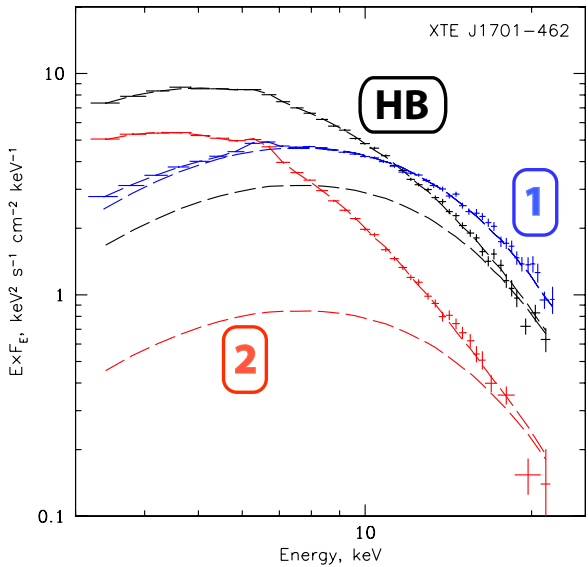}
\caption{Three typical spectra of XTE J1701$-$462 at different parts of its colour-intensity diagram along with their best-fitting models. 
The spectra are labelled with numbers corresponding to the positions shown in Fig.~\ref{fig:cid}. 
The model consists of the multicolour blackbody component ({\sc diskbb}), the BL/SL model 
and a Gaussian line at 6.4 keV. 
The dashed curves show the shape of the BL/SL component (the same shape for all three spectra) taken from the fit to 
the Fourier frequency resolved spectra (see Fig.~\ref{fig:freqspectra}). Other components are not shown for clarity.
}
\label{fig:threedecomp}
\end{figure}

The resulting Fourier frequency resolved spectra are shown in Fig.~\ref{fig:freqspectra}. It is clearly seen that similarly to the case of GX~340+0, considered in details by \citet{gilfanov03},  all frequency resolved spectra at Fourier frequencies above 1 Hz have the same spectral shape. 
Moreover, this spectral shape is almost identical to those of the Fourier frequency resolved energy spectra of all sources  analysed in \citet{gilfanov03} and \citet{revnivtsev06} and interpreted as spectra of the BL/SL on the NS surface.

This spectrum can be adequately described by a simple analytical formula (see Fig.~\ref{fig:freqspectra}) 
\begin{equation} \label{eq:sl_spectrum}
\frac{dN}{dE} \propto E^{-\Gamma} \exp(-E/E_{\rm c}), 
\end{equation}
with the photon index $\Gamma\approx$0.0--0.1 and the cutoff energy of $E_{\rm c}\approx$3.5--3.8~keV.
The best-fitting spectral parameters in all frequency intervals are within these ranges
with the errors approximately equal to half of the interval width.
 A more physically motivated spectral approximation of the frequency resolved spectra by a simple model of saturated Comptonization ({\sc comptt} model in {\sc xspec}  package) gives the temperature of the seed photons 
 $kT_{\rm s}\sim 0.8-1$~keV, temperature of Comptonizing electrons $kT_{\rm e}\sim2.8-3.0$~keV 
 and their optical depth $\tau\sim 7-8$. 
 These values are similar to those obtained for other luminous NS binaries in their soft state \citep{revnivtsev06}.

The resulting decomposition of the energy spectra of XTE J1701$-$462 
at different locations of the colour--intensity diagram, denoted as "1", "2" and "HB" and shown in Fig.~\ref{fig:cid}, 
into the accretion disc and the BL/SL components is shown in Fig.~\ref{fig:threedecomp}. 
It can be seen that the quality of this simple fit is reasonable 
($\chi^2\sim 38-40$ for 40 dof) and the BL/SL component dominates at energies above 10--15 keV.

\subsection{Is this the Eddington-limited emission?}

The key feature of the SL at the NS surface is the importance of the balance
between  the gravity and  the sum of the centrifugal and radiation pressure forces  
\citep{IS99}. The contributions  of the centrifugal and radiation pressure forces 
change over the SL latitude, from the centrifugally-supported region at the 
NS equator to the radiation-pressure-supported region at the highest SL latitude.
 In the optically thick regime, it  means that the maximum effective temperature and, therefore, the 
 maximum colour temperature of the emission is set mainly by the gravity at the NS surface \citep{goldman79,marshall82,revnivtsev06}. 
 If the effective temperature of the emission exceeds the corresponding Eddington value (set by the gravity), 
 the surface layers should be blown away by the  radiation pressure force. 

This is exactly what is observed in some powerful thermonuclear (type I) X-ray bursts, when  at high luminosity level the radius of the NS photosphere starts to expand. It means that the colour temperatures of the NS photosphere at the beginning of the photospheric radius expansion (PRE) phase should be close to that of the radiation-pressure dominated BL/SL. The case of XTE J1701$-$461 provides us a possibility to check this, because  this source  demonstrated the type I bursts \citep{lin09b}. 

\begin{figure}
\includegraphics[width=0.95\columnwidth,bb=32 182 570 700,clip]{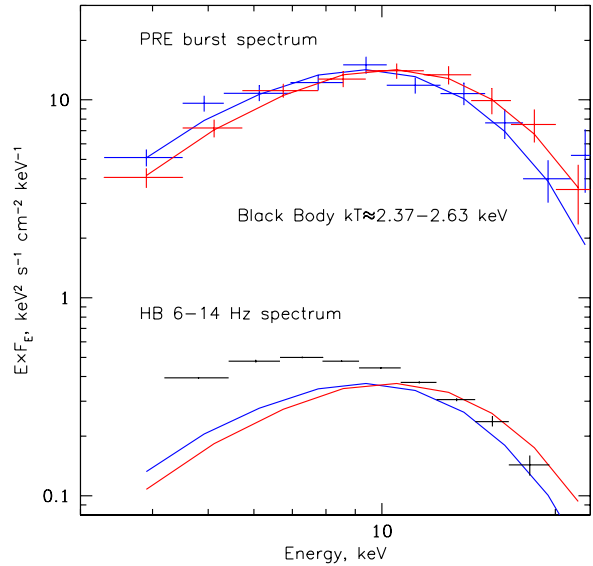}
\caption{The spectrum of XTE J1701$-$462 just before the PRE phase during the X-ray burst  
(blue crosses), the spectrum of the source after the touchdown (red crosses)
 and the Fourier frequency resolved energy spectrum of the source (black crosses). 
The solid curves show the shapes of the blackbody spectra with temperatures $kT_{\rm bb}=2.37$ (blue) 
and  $2.63$ keV (red). The lower pair of  curves are the same curves as above, 
scaled down to match the level of the frequency resolved spectrum at energies 12--20 keV.
}
\label{fig:sl_and_burst}
\end{figure}

In Fig.~\ref{fig:sl_and_burst}, we present three spectra: the spectrum of the source just before the start of the 
PRE phase of the X-ray bursts observed on 2007 July 20 
\citep[see][for the detailed analysis of the bursts]{lin09b}, the spectrum after the photosphere has settled 
down to the NS surface and the Fourier frequency resolved spectrum at horizontal branch of the Z-diagram of  XTE J1701$-$462, 
which is essentially the spectrum of the SL. We see that 
the colour temperature of the burst spectra are similar to the maximal colour temperature of the SL spectrum. This, therefore, supports our conclusion that the maximal colour temperature of the SL is set mainly by the NS  gravity.
The spectrum of the SL is wider, likely due to some distribution of effective temperatures over the SL surface. 
This might be caused by a variation of the role of the centrifugal force  
over the  NS latitude that causes changes in the vertical structure of the SL \citep[see][]{IS99}.

\subsection{Does SL spectrum vary with luminosity?}

One of the very distinct feature of the colour-intensity diagram of XTE J1701$-$462 (see Fig.~\ref{fig:cid}) 
is the maximum level of the hard colour during its soft state (i.e. with fluxes above $\sim 10^{-9}$ erg s$^{-1}$ cm$^{-2}$). The scatter plot created by behaviour of the source in its Z-state (more specifically, on the flaring branch of the Sco-like Z-state, see \citealt{lin09a,homan10} for details of the classification) shows ``flaring'' tracks, which end up at the level of hardness $\approx0.95-1.05$. This hardness very closely corresponds to the hardness of the BL/SL spectrum, as determined from the Fourier frequency resolved spectra. A more detailed comparison of the Fourier frequency resolved spectrum and the hardest energy spectra at the top of the flaring branch shows that during these moments the BL/SL is the dominant contributor to the total emission of the source.

\begin{figure}
\includegraphics[width=0.95\columnwidth,bb=32 182 570 720,clip]{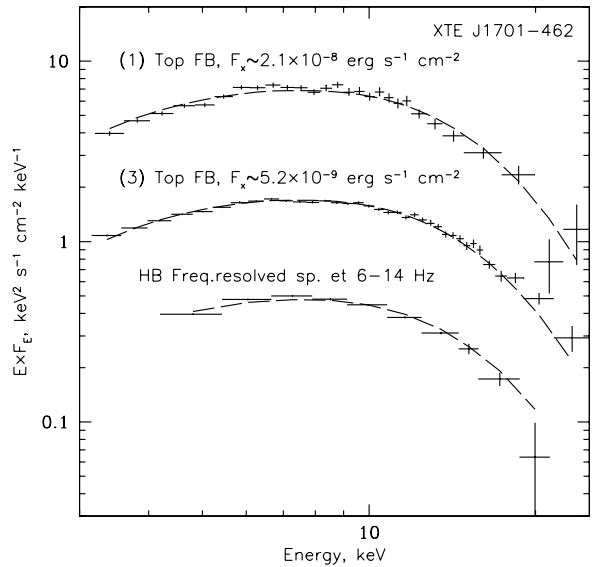}
\caption{Three energy spectra of XTE J1701$-$462. The two top spectra have been collected at the upper parts of the ``flaring'' branch at the colour-intensity diagram of the source, where the hard colour has the value above 0.95 (numbers denote the location of source at the colour-intensity diagram on Fig.~\ref{fig:cid}). 
The lowest spectrum is the spectrum at Fourier frequencies 6--14 Hz and represents the spectrum of the BL/SL.
}
\label{fig:frsp_and_twoFB}
\end{figure}

In order to demonstrate this fact, we present three spectra at Fig.~\ref{fig:frsp_and_twoFB}. The top two spectra have been collected at the top parts of tracks on the colour-intensity diagram, which have hard colour above 0.96, but have very different fluxes. The upper spectrum has the absorption corrected flux $F_{\rm x}\sim2.1\times10^{-8}$ erg s$^{-1}$ cm$^{-2}$ (position is denoted as "1" on Fig.~\ref{fig:cid}) and the lower one -- $F_{\rm x}\sim5.2\times10^{-9}$ erg s$^{-1}$ cm$^{-2}$ (position is denoted as "3" on Fig.~\ref{fig:cid}). 
The lowest spectrum is the Fourier frequency resolved spectrum taken at 6--14 Hz.
It is clear that the spectra have similar (though not completely identical) spectral shapes and could be adequately described within the energy range 3-20 keV by either a power law model with the exponential cutoff $dN/dE\propto E^{-0.05}\exp(-E/3.8 \ \mbox{keV})$, or by a model of saturated Comptonization {\sc comptt} (with temperature of the seed photons 
$kT_{\rm s}=0.73$ keV, temperature of the Comptonizing electrons $kT_{\rm e}=2.69$ keV and their optical depth $\tau=7.64$).
The curve shows the  {\sc  comptt} spectral model scaled to match the different levels of the spectra (the quality of the fits by the analytical model of a power law with the exponential cutoff is similar).
Similarity of these spectra allows us to conclude that at the top parts of the flaring tracks we see only emission of the SL.

\begin{figure}
\includegraphics[width=0.95\columnwidth,bb=32 182 570 720,clip]{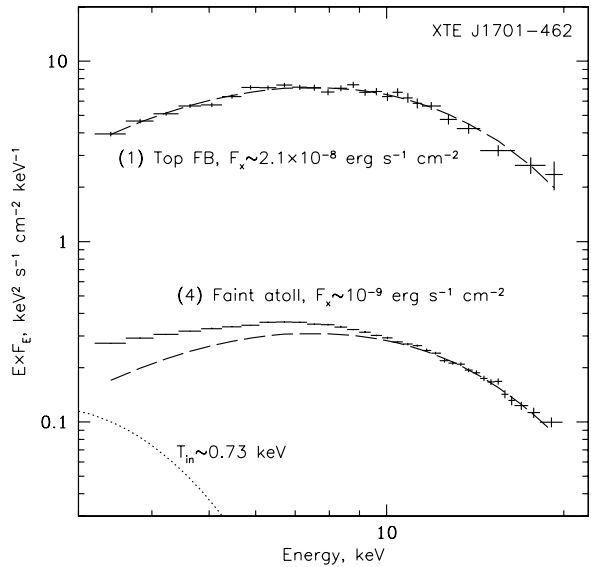}
\caption{The energy spectra of XTE J1701$-$462  collected at different time periods. The top spectrum is the same as in Fig.~\ref{fig:frsp_and_twoFB}, the lower one is collected during the faintest sources fluxes, at which the source is still in its soft state (fluxes around $10^{-9}$ erg s$^{-1}$ cm$^{-2}$). 
The spectra are labeled by the numbers corresponding to the location of the source at the 
colour--intensity diagram as shown in Fig.~\ref{fig:cid}. 
The dashed curves show the contribution of the BL/SL component and the dotted curve is the accretion disc component.}
\label{fig:bright_and_faint}
\end{figure}

At fainter fluxes (e.g. at position denoted as "4" on Fig.~\ref{fig:cid}), the source does not demonstrate the flaring behaviour anymore and therefore we cannot study the BL/SL emission in its ``clear'' state. 
However, as the BL/SL emission still dominates at high energies, we can look at the spectral shape there. 
In Fig.~\ref{fig:bright_and_faint}, we show that the colour temperature of the emission measured in the energy band 12--20 keV is perfectly compatible with that of the BL/SL at all larger fluxes. 
The excess clearly visible at energies below 10--12 keV can be attributed to the emission of the accretion disc.

There are two important conclusions from these spectral models:
\begin{itemize}
\item At the top  part of the flaring branch (with the largest hard colour), the BL/SL is virtually the only contributor to the total emission. 
 The accretion disc component gives a negligible contribution to the observed spectrum. 
\item The BL/SL spectrum does not change its spectral shape over at least a factor of 20 variations in its flux. The ``Wien tail'' colour temperature 
of the SL emission stays constant at the level of 2.4--2.6 keV. 
\end{itemize}

\section{Summary}


We have analysed a complete dataset of observations of a unique NS transient XTE~J1701$-$462, which has demonstrated for the first time a 
large variety of patterns of spectral-timing behaviour.
A large span of the source luminosities has allowed us to check the stability of the BL/SL spectrum.  
Our main results can be summarised as follows:
\begin{enumerate}
\item The decomposition of the energy spectra of  XTE J1701$-$462 into constituent components cannot be unambiguously done with the help of the spectral fitting only.
The assumptions about the spectral shape of the components play a crucial role in the resulting decomposition. 
\item We have demonstrated that the Fourier frequency resolved energy spectra of the source taken at frequencies 
above 1 Hz is adequately represented by one spectral component. 
We have argued that this spectral component originates from the BL/SL on the NS surface.
The shape of this component is very similar to those extracted for a set of other sources, such as GX~340+0, 4U~1608$-$52, GX~17+2, 
Cyg~X-2 and 4U~1820$-$30  \citep{revnivtsev06}. 
\item We have demonstrated that the maximum colour temperature of the BL/SL spectrum does not vary over a factor of more than 20 variation of its flux.
These findings strongly support the theoretical model of the BL/SL of \citet{IS99}, which states that the maximum colour temperature of the BL/SL is set by NS gravity. 
Elaboration of this model will provide solid ground for accurate measurements of the NS masses and radii from X-ray observations.
\end{enumerate}

\section*{Acknowledgements}

MGR acknowledges the support from the grant NSh-5603.2012.2 of the 
program P19 of the Presidium of the Russian Academy of Sciences (RAS) and the 
program OFN17 of the Division of Physical Sciences of the RAS.
The work  of VFS is supported by the German Research Foundation
(DFG) grant SFB/Transregio 7 "Gravitational Wave Astronomy" and the 
Russian Foundation for Basic Research (grant 12-02-97006-r-povolzhe-a). 
This work was also supported by 
the Academy of  Finland exchange program  grants 259284 and 259490 (VFS, JP), 
and by the Jenny and Antti Wihuri foundation (VFS).


\label{lastpage}

\end{document}